# Imaging Cooper Pairing of Heavy Fermions in CeCoIn$_5$

M. P. Allan[1,2,3,4§], F. Massee[1,2§], D. K. Morr[5], J. van Dyke[5], A.W. Rost[2,3], A. P. Mackenzie[3,6], C. Petrovic[1] and J. C. Davis[1,2,3,7]

[1] CMPMS Department, Brookhaven National Laboratory, Upton, NY 11973, USA
[2] LASSP, Department of Physics, Cornell University, Ithaca, NY 14853, USA.
[3] School of Physics and Astronomy, University of St Andrews, St Andrews, Fife KY16 9SS, UK.
[4] Department of Physics, ETH Zurich, CH-8093 Zurich, Switzerland
[5] Department of Physics, University of Illinois at Chicago, Chicago, IL 60607, USA
[6] Max-Planck Institute for Chemical Physics of Solids, D-01187 Dresden, Germany
[7] Kavli Institute at Cornell for Nanoscale Science, Cornell University, Ithaca, NY 14853, USA.
§ These authors contributed equally to this research project.

**The Cooper pairing mechanism of heavy-fermion superconductors[1-4], while long hypothesized as due to spin fluctuations[5-7], has not been determined. It is the momentum space ($k$-space) structure of the superconducting energy gap $\Delta(k)$ that encodes specifics of this pairing mechanism. However, because the energy scales are so low, it has not been possible to directly measure $\Delta(k)$ for any heavy-fermion superconductor. Bogoliubov quasiparticle interference (QPI) imaging[8-10], a proven technique for measuring the energy gaps of high-T$_c$ superconductors[11-13], has recently been proposed[14] as a new method to measure $\Delta(k)$ in heavy-fermion superconductors, specifically CeCoIn$_5$ [15]. By implementing this method, we immediately detect a superconducting energy gap whose nodes are oriented along $k||(\pm1,\pm1)\pi/a_0$ directions[16-19]. Moreover, we determine the complete $k$-space structure of the $\Delta(k)$ of a heavy-fermion superconductor. For CeCoIn$_5$, this novel information includes: the complex band structure and Fermi surface of the hybridized heavy bands, the fact that highest magnitude $\Delta(k)$ opens on a high-$k$ band so that gap nodes occur at quite unanticipated $k$-space locations, and that the Bogoliubov quasiparticle interference patterns are most consistent with d$_{x^2-y^2}$ gap symmetry. The availability of such quantitative heavy band- and gap-structure data will be critical in identifying the microscopic mechanism of heavy fermion superconductivity in this material, and perhaps in general.**

The heavy-fermion superconductor CeCoIn$_5$ (Ref. 15) has a crystal unit cell with $a=b=4.6$Å, $c=7.51$Å as shown schematically in Fig. 1a, and a superconducting critical temperature $T_c=2.3$K. If antiferromagnetically ordered, the Ce$^{3+}$ atoms would exhibit local magnetic moments $\mu=0.15\mu_B$.[20] While that state does not exist in the pure compound studied here, antiferromagnetic spin fluctuations do persist [21]. In the superconducting phase, the Cooper pairs are spin singlets[22,23] so that an even parity $\Delta(\mathbf{k})$ is required. Magnetic field-angle dependence of thermal conductivity[17] and specific heat[19] are interpreted as evidence of energy-gap nodes $|\Delta(\mathbf{k})|=0$ for momentum space directions $\mathbf{k}||(\pm 1,\pm 1)\pi/a_0$. However, a fully detailed knowledge of the $\mathbf{k}$-space structure of $\Delta(\mathbf{k})$ is required to understand the microscopic Cooper pairing mechanism of heavy-fermions. This cannot be achieved using such indirect methods, or by using photoemission because the energy resolution required is $\delta E<100$ μeV. Motivated thus, high resolution Bogoliubov quasiparticle scattering interference imaging has recently been mooted[14] as a promising approach for determining $\Delta(\mathbf{k})$ of heavy-fermion superconductors, specifically for CeCoIn$_5$.

There are three elements of $\mathbf{k}$-space electronic structure expected in a generic heavy-fermion superconductor[4]. First, the high temperature state consists of a conventional (light) electronic band indicated schematically by the dashed curve in Fig. 1b that coexists with localized $f$-electron states on each magnetic atom. At lower temperatures, hybridization between this light band and the $f$-electron states results in its splitting into two new heavy bands as shown schematically by the solid blue lines in Fig. 1b. The right panel shows how the resulting very flat bands generate a greatly enhanced density-of-electronic-states $N(E)$ within a few meV of $E_F$ - hence the 'heavy' effects seen in thermodynamic studies. At least one of these heavy bands crosses $E_F$ at the new Fermi wavevector $\mathbf{k}_F^H$ as shown within the green box in Fig. 1b. It is in this region of $\mathbf{k}$-space that, at even lower temperatures, the heavy quasiparticles are hypothesized to bind into heavy

Cooper pairs. An energetically particle-hole symmetric superconducting energy gap $\Delta(\boldsymbol{k})$, probably of an unconventional nature[1-7], is then expected to open in the heavy quasiparticle spectrum at the Fermi surface, as shown schematically in Fig. 1c. The right panel shows the expected additional changes in $N(E)$ for a nodal $\Delta(\boldsymbol{k})$.

To search for this sequence of phenomena, we use pure $CeCoIn_5$ samples inserted into the cryogenic ultra high vacuum of a $^3$He-refrigerator-based spectroscopic imaging scanning tunneling microscope (SI-STM), and mechanically cleaved therein. Atomically flat *a-b* surfaces are achieved; a typical resulting topograph with the Ce or Co atomic lattice $a_0$=4.6Å visible is shown in Fig. 1d. On all such surfaces, the density-of-states N($E$) is determined from the spatially averaged differential tunneling conductance <d$I$/d$V$($E$=eV)>$\propto$ N($E$) measured far from impurity atoms. While the basic N($E$) of the unhybridized 'light' bands is measured over the range $|E|\leq$200 meV (Fig. 1e), the complex scattering interference features associated with the heavy band structure are only visible within the range -4meV<$E$<12meV. Vertical arrows in Fig. 1f then indicate the limits of the hybridization gap $E_h$ for $CeCoIn_5$, as determined directly from the heavy band scattering interference analysis in Fig. 3 below.

Upon entering the superconducting phase, $\bar{N}(E)$ develops an energy gap with maximum value $|\Delta_{max}|$=550±50 μeV, a V-shaped $\bar{N}(E)\propto E$ that is the signature of a nodal[24,25] $\Delta(\boldsymbol{k})$, and a finite[24-26] $N(E=0)$ all as shown in Fig. 1g. Figure 2a shows a typical example of atomically resolved images g($\boldsymbol{r},E$) ≡ d$I$/d$V$($\boldsymbol{r},E$=eV) measured within the superconducting gap at $E$=250 μeV, and acquired in 32 nm x 32nm field of view (FOV). The superconducting gapmap $\Delta_{pp}(\boldsymbol{r})$ in the same FOV (Fig. 2b) reveals the electronic homogeneity of this material. In Fig. 2c we show an image of the $CeCoIn_5$ Abrikosov vortex array acquired at $T$=250mK, $B$=3T in a larger FOV; its shape and orientation are in excellent agreement with small-angle neutron scattering studies[27]. As all these phenomena disappear at the superconducting T$_c$ observed in bulk measurements, the $|\Delta|$=550±50 μeV energy gap with V-shaped $N(E)$ is definitely that of the superconductor. Determination of the $\boldsymbol{k}$-space structure of $\Delta(\boldsymbol{k})$ for CeCoIn$_5$ is then the main focus of this paper.

To proceed, we image the differential conductance g($\boldsymbol{r}$,$E$) with atomic resolution and register, and then determine g($\boldsymbol{q}$,$E$), the square root of the power spectral density Fourier transform of each image. To investigate both how the light bands transform to hybridized heavy fermion states, and how superconductivity then emerges, we measure these datasets on three distinct energy scales, each of about an order of magnitude smaller energy range than the previous one, as described in the supplementary information (SI) Section I. These data are used to evaluate elements of $\boldsymbol{k}$-space electronic structure, based on the fact that elastic scattering of electrons with momentum -$\boldsymbol{k}$($E$) to +$\boldsymbol{k}$($E$) generates interference patterns occurring as maxima at $\boldsymbol{q}$(E)=2$\boldsymbol{k}$($E$) in g($\boldsymbol{q}$, $E$), an effect recently revealed[28,29] to exist even when hybridization generates heavy-fermion bands. In SI Section I we show the measured g($\boldsymbol{q}$,$E$) at $T$=1.2K, for -100meV<$E$<30meV focusing on the light unhybridized electronic structure. Here the maximum intensity features move slowly and smoothly to smaller |$\boldsymbol{q}$|-radii with increasing E thereby revealing a light and simple tetragonal band (SI Section I $g$($\boldsymbol{q}$,$E$) movie S1).

Drastic departures from this simple phenomenology are found to occur only within the energy range -4meV<E<12meV. In Fig. 3a-e we next show the measured $g$($\boldsymbol{q}$,$E$) at $T$=250mK (and thus energy resolution $\delta E$≈3.5k$_B$$T$≈75μeV) within this range (SI, Section I). The onset of hybridization is detected as a sudden transformation of the previously unchanging structure of g($\boldsymbol{q}$,$E$) occurring at $E$≈-4 meV (Fig. 3b) followed by a rapid evolution of the maximum intensity features (indicated by circles and arrows Fig. 3c) towards smaller |$\boldsymbol{q}$|-radius interference patterns. Then, in Fig. 3c we see that an abrupt jump to a larger |$\boldsymbol{q}$|-radius occurs, followed by a second rapid diminution of interference pattern |$\boldsymbol{q}$|-radii in Figs 3c-e (complete phenomena in SI $g$($\boldsymbol{q}$,$E$) movie S1). These are all the expected QPI signatures of the appearance of hybridized heavy-fermion bands[28,29]. Thus, for



CeCoIn$_5$ this approach reveals how the light conduction band is split into two heavy bands within the hybridization gap -4<$E_h$<12 meV. To see this directly, we show in Fig. 3f,g the measured evolution of the maxima in g($q$,$E$) for two directions in $q$-space. The light band (grey dots) begins to deviate near -4meV towards the lower heavy band which crosses $E_F$ at smaller |$q$|=2|$k_F^H$|, and evolves quickly to even smaller |$q$| (blue dots). Within a few meV above $E_F$, the interference patterns jump to a much larger |$q$| and then evolve (blue dots) back towards the light band (grey dots) which they rejoin near +12meV. This heavy band actually crosses below E=0 at high $k$, producing an electron-like Fermi surface whose intra-band scattering interference generates interference patterns at low $q$ (blue dots E<0 as |$q$|→0). These data (Fig. 3a-e, SI Section I), and the extracted dispersions (Fig. 3f,g) are next used to determine details of the heavy fermion band structure.

In general for a complex and multi-band $k$-space structure, achieving a deterministic inversion procedure from g($q$,$E$) data to the complete band structure can be challenging[13]. Here, comparison of the predicted scattering interference dispersions |$q$($E$)| from a specific model of the heavy bands described in SI Section II and Fig. 3h, with the experimental |$q$($E$)| data within the hybridization range $E_h$, reveals good agreement (Supplementary Figs S2, S3). The critical elements in our model that lead to this agreement are the nearly parallel sections of the light band contours-of-constant-energy and the hybridization with a specifically shaped f-band. Some of these elements can equally be found in the model developed by Ref. 14, which exhibits a very similar Fermi surface. However, our model concentrates on best emulation of the key empirical phenomena of heavy QPI. On this basis, the g($q$,$E$) in Fig. 3a-g are used to motivate the detailed $k$-space model for the heavy bands of CeCoIn$_5$ as shown in Fig. 3h (SI Section II) . Here, within the range $E_h$, a light-hole like band centered around Γ (or equivalently M) hybridizes with a localized f-electron band (SI Section II). The resulting lower heavy band β has a simple Fermi surface and closes quickly above $E_F$, while the upper heavy band α is highly anisotropic with a complex Fermi surface as it crosses below $E_F$ making it



effectively electron like. Our model indicates the possibility of a small dimple that crosses back above $E_F$ but this is in no way critical to the subsequent analysis. The Fermi surfaces are shown as solid lines on the $E$=0 planes of Fig. 3h.

To explore the superconductivity on the heavy bands in Fig. 3h, Figs 4a-e show the measured g($q$,$E$) at $T$=250mK $|E|$<300μeV , within the superconducting energy gap. Here we see extremely rapid evolution in $g(q,E)$ over energies of a few 100 μeV, and the appearance of a four-fold symmetric "nodal" $g(q,E)$ structure as $E\rightarrow0$. Clearly, this $g(q,E=0)$ exhibits far more complexity than expected for a single-band nodal superconducting energy gap[8,9,10,13]. To explore these phenomena we carry out Bogoliubov QPI simulations based upon the two heavy bands, $\alpha$ and $\beta$ (Fig. 4h,4k) but now specifying their superconducting energy gaps $\Delta_\alpha(k)$ and $\Delta_\beta(k)$, whose derivation is discussed below. Here the inter-nodal scattering wavevectors for the $\alpha$ band (colored arrows Fig. 4k) are demonstrably consistent with the measured inter-nodal scattering vectors in g($q$,$E$=0) data, while the equivalent internodal signatures are completely absent for the $\beta$ band. As specific heat data show that all the main bands in CeCoIn$_5$ are gapped at lowest temperatures[30], this suggests that the gap on the $\beta$ band is too small to be detected at T~250mK. What our data do indicate is that the primary gap of CeCoIn$_5$ actually occurs on the high-$k$ $\alpha$ band with lines of gap-nodes along the $k$=(0,0)→($\pm\pi,\pm\pi$)/a$_0$ directions, so that the actual gap nodes in CeCoIn$_5$ occur at unanticipated $k$-space locations (Fig. 4k).

Next we consider a detailed comparison of the measured g($q$,$E$) data for $|E|$≤550 μeV at $T$~250 mK with theoretical simulations of Bogoliubov QPI in $g(q,E)$ using the $\alpha,\beta$ Fermi surfaces described in Figs 3h,4k. The simulations have been carried out using various symmetries for the superconducting energy gaps. Our approximate model with $\Delta_\beta(\theta_k)$=0 and a d$_{x^2-y^2}$ symmetry gap $\Delta_\alpha(\theta_k)$=$ACos(2\theta_k)$ with $A$=550±50 μeV yields a set of simulated g($q$,E)  that are far more consistent with the experimental data than any of the other models we have considered (SI Section III). For comparison, a direct experimental estimation of $|\Delta(\theta_k)|$ can be



achieved by using $g(E$-$q)$ data in a procedure largely independent of the Fermi surface details. To obtain the angle $\theta_q$ dependence of the energy gap that opens at $T_c$, we integrate the total spectral weight g($q$,E) within a given $|\delta q|$ range containing the Fermi surface, with lowest $|q|$ large enough to exclude effects of heterogeneity and largest $|q|$ small enough to exclude the Bragg peaks. A clear gap $\Delta(\theta_q)$ is observed to open in this integral of g($q$,E) upon passing below $T_c$, as demonstrated in SI Section IV. In Fig. 4l we plot the measured energy gap $|\Delta(\theta_q)|$ from this technique (red dots) along with $\Delta_\alpha(\theta_k)=ACos(2\theta_k)$; $A$=550 µeV as a (solid line). Their agreement provides strong independent motivation for our gap structure model (Fig. 4k). A final stimulating observation revealed here is that the departures in the $\Delta(\theta_q)$ data at higher energy from the simple $\Delta_\alpha(\theta_k)=ACos(2\theta_k)$, might be expected if high $q$ scattering between these locations on the $\alpha$ band is involved in the Cooper pairing mechanism.

Overall, these data represent a direct measurement of the $k$-space structure of the superconducting energy gaps $\Delta.(k)$ for a heavy-fermion superconductor. They reveal a wealth of previously unknown information on $\Delta(k)$ of CeCoIn$_5$ including: (i) the complex Fermi surface of the hybridized heavy bands (Fig.s 3h,4k); (ii) the spectroscopic signature of four nodal lines in $|\Delta(k)|$ oriented along $k=(\pm 1,\pm 1)\pi/a_0$ or Ce-Ce directions[16-22]; (iii) that the dominant $\Delta(k)$ opens on the $\alpha$ heavy band at high $k$ (Fig. 4k); (iv) the resulting unanticipated $k$-space locations of the gap nodes (Fig. 4k) ; (v) that the Bogoliubov QPI patterns are most consistent with $d_{x^2-y^2}$ gap symmetry, and (vi) evidence for a departure in $\Delta(k)$ from a simple $Cos(2\theta_k)$ dependence on the $\alpha$ band (Fig. 4l). These highly specific multi-band $\Delta.(k)$ data provide the information critical for determination of the microscopic mechanism of heavy fermion superconductivity in CeCoIn$_5$.

## Materials and Methods



High quality CeCoIn$_5$ single crystals were grown at BNL details in Ref. 15. Magnetization measurements prior to sample insertion into the STM show a sharp transition with T$_c$ = 2.1 K. The samples were mechanically cleaved in cryogenic ultrahigh vacuum at $T\sim$10 K and directly inserted into the STM head at 4.2 K. Etched atomically sharp and stable tungsten tips with energy independent density of states are used. Differential conductance measurements throughout used a standard lock-in amplifier. See Supplementary Information for additional details on data treatment and extraction.

**Figure Legends**

**Fig. 1 Anticipated Electronic Structure of a Heavy-fermion Superconductor**

a    Schematic representation of crystal unit cell of CeCoIn$_5$.

b    Schematic of typical evolution of *k*-space electronic structure observed as hybridization splits the light band into two heavy bands[28], and the consequential effects on the density of states *N(E)*.

c    Schematic of expected evolution of *k*-space electronic structure as the superconducting energy gap appears (presumably) on one of the new heavy bands. The right-hand panel shows expected changes in the *N(E)* due to heavy-fermion Cooper pairing, here simulated for a d-wave symmetry energy gap.

d    Topographic image of termination surface of cryo-cleaved CeCoIn$_5$ used in this study.

e    Average differential conductance spectra g(*E*) in the energy range of light band(s) |*E*|≤200meV, measured using the lock-in technique with a bias modulation of 5meV so that any finer energetic features are unresolvable. Data in f and g below are acquired with decreasing junction resistance compared to that in e.

f    Measured average differential conductance spectra in the energy range spanning the hybridization gap ∼-4meV<*E*<12meV. The hybridization gap E$_h$



between vertical arrows is determined directly from heavy-quasiparticle scattering interference (Fig. 3), measured with bias modulation 1.5meV so that any finer energetic features, e.g. the superconducting energy gap, are unresolvable.

**g**  Measured differential conductance spectra in the energy range spanning the superconducting gap $|E| \leq 600$ μeV, measured with a bias modulation of 70μeV and a thermal energy resolution of 75μeV.

**Figure 2 Imaging Superconducting Gapmap and Vortex Lattice of CeCoIn$_5$**

**a**  Typical example of $g(\mathbf{r},E)$ measured below the superconducting gap edge $|\Delta|=550$ μeV and acquired in the 32 nm x 32nm field of view (FOV).

**b**  Superconducting gapmap $\Delta_{pp}(\mathbf{r})$ measured between the particle-hole symmetric peaks in g($\mathbf{r}$,E) taken in same FOV as a. The homogeneity of the gap structure away from impurities is as expected in these pure materials. The inset shows a typical spectrum with arrows denoting the maximal gap, $\Delta_{pp}$.

**c**  Image of CeCoIn$_5$ $\Phi$=h/2e Abrikosov vortex array at B=3T by measuring g($\mathbf{r}$,E=0,3T)-g($\mathbf{r}$,E=$\Delta_{pp}$,3T). The lattice is consistent with the square lattice reported by neutron scattering experiments[27], taking into account the small field drift.

**Figure 3 Heavy-fermion Bands and Fermi Surfaces in CeCoIn$_5$**

**a-e**  Measured g($\mathbf{q}$,E) at $T$=250 mK and $\delta E \sim 75$ μeV, within the heavy-fermion forming hybridization window -4meV<$E_h$<12meV. The numbered arrows indicate locations of maxima in g(q,E) who dispersion is identified using similarly numbered arrows in f, g.

**f,g**  Measured evolution of the light band scattering interference dispersion $|\mathbf{q}(E)|$ (grey circles) and its transition to two heavy bands (blue circles) each with a distinct $|\mathbf{q}(E)|$. Some points are fitted on g($\mathbf{q}$,E=const) layers, see while others are fitted from g($|\mathbf{q}|$,E) cuts (SI section V).



**h**  Momentum-space model for the hybridization induced heavy bands and Fermi surfaces of CeCoIn$_5$. Detailed parameterization is given in SI Section V. At the center of the upper half of this panel we see the light band (grey) closing at the center. As $E$=0 is approached from above, the upper heavy band (blue) diverges from the light band and begins to disperse very rapidly outwards and crosses $E$=0 at high $k$. The lower half of this panel shows the light band (grey) approach E=0 from below, beginning to diverge rapidly towards low $k$ as it crosses $E$=0 (blue) and then closing just above $E$=0. The characteristic Fermi surface areas deduced from the data/model for the heavy bands shown in Fig. 3 are in reasonable agreement to those found in quantum oscillation studies in CeCoIn$_5$[31]. Furthermore, data on the CeCoIn$_5$ $k$-space structure measured using both SI-STM[32] and ARPES[33,34] at $T$~20K (energy resolution $\delta E \approx 3.5 k_B T \approx 5$meV) are not inconsistent with the far higher precision *($\delta E \approx 3.5 k_B T \approx 75 \mu eV$)* heavy-band determinations herein.

**Figure 4 Momentum-space Superconducting Energy Gap $\Delta(k)$ of CeCoIn$_5$**

**a-e**  Measured g($q,E$) at $T$=250 mK and $\delta E$~75 μeV, within the heavy-fermion superconductivity energy window -550μeV<$E$<550μeV.

**f-j**  Bogoliubov QPI simulations of g($q,E$) (SI Section II) on the two bands as shown in Fig 3h, 4k, and for $\Delta_\beta(\theta_k)$=0 and $\Delta_\alpha(\theta_k)$=$A$cos(2$\theta_k$) with $A$=550 μeV.

**k**  Fermi surfaces and energy gaps of CeCoIn$_5$ modeled using heavy QPI in Fig. 3 (SI Section II). The superconducting energy gaps $\Delta_i(k)$ used to achieve the most successful BQPI simulations are shown in red. The internodal scattering vectors consistent with the data are show as solid arrows while the Friedel oscillation wavevectors of the ungapped (at 250mK) Fermi surface regions are shown as dashed lines (details in SI Section II).

**l**  Measured $\Delta(\theta_q)$ using techniques as described in text (SI Section IV) and its comparison with the simplest multi-band gap structure $\Delta_\beta(\theta_k)$=0 and $\Delta_\alpha(\theta_k)$=$A$cos(2$\theta_k$); $A$=550 μeV that we find to be consistent with all the



Bogoliubov g($q$,$E$) data herein. Arrows identify the strong departures form this simple gap function.

**Acknowledgements:** We are particularly grateful to I. Ermin, J. E. Hoffman and A.R. Schmidt for advice and discussions. We acknowledge and thank A. Akbari, M. Aprili, M. H. Fischer, M. Hamidian, E.-A. Kim, S. A. Kivelson, M. Norman, J.P. Reid, D.-H. Lee, M. Norman, D.J. Scalapino, and K. Shen for helpful discussions , advice and communications. Supported by US DOE under contract number DEAC02-98CH10886 (J.C.D. and C.P.) and under Award No. DE-FG02-05ER46225 (DKM, JvD); by the UK EPSRC; APM acknowledges support of a Royal Society-Wolfson Award.

**Author Contributions:** F.M. and M.P.A. carried out SI-STM experiments plus the data preparation; C.P. synthesized and characterized the samples; D.K.M. and J.vD. developed the band and gap structure models; J.C.D. supervised the project and wrote the paper with key contributions from F.M, M.P.A. D.K.M. and A.P.M. The manuscript reflects contribution and ideas of all authors.

* To whom correspondence should be addressed: jcseamusdavis@gmail.com

**Fig 1**

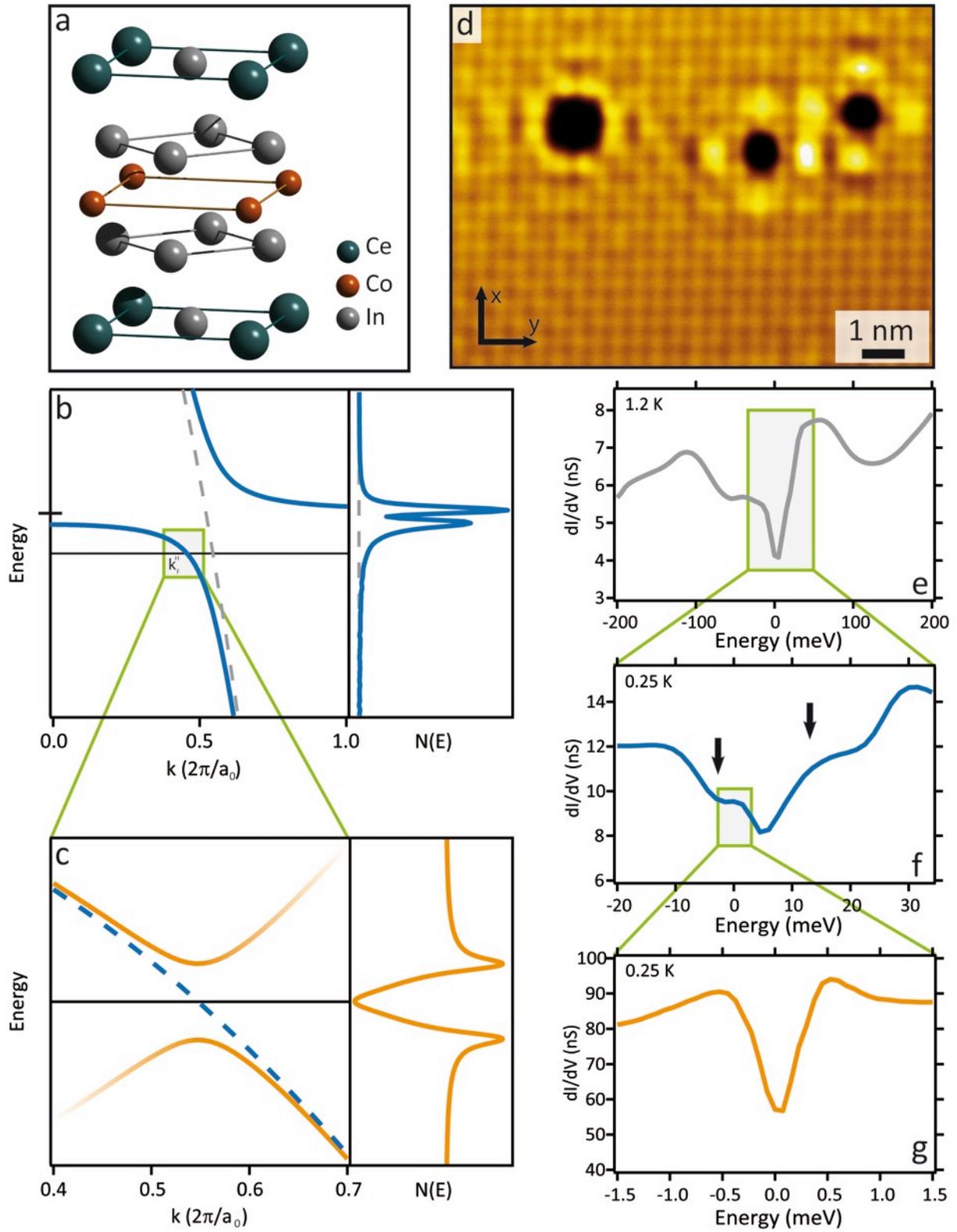

**Fig 2**

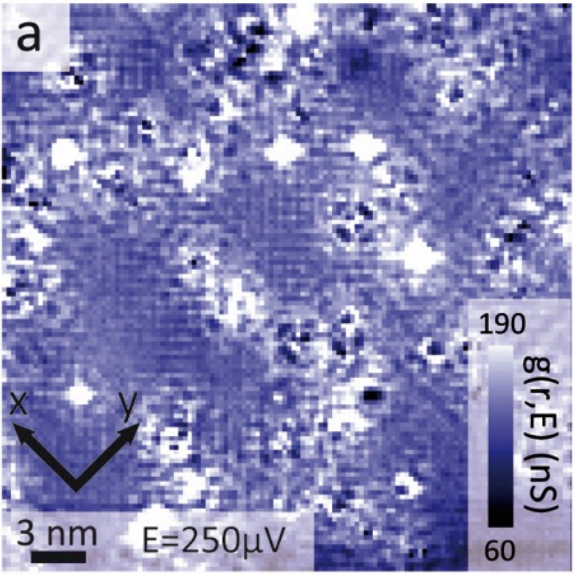
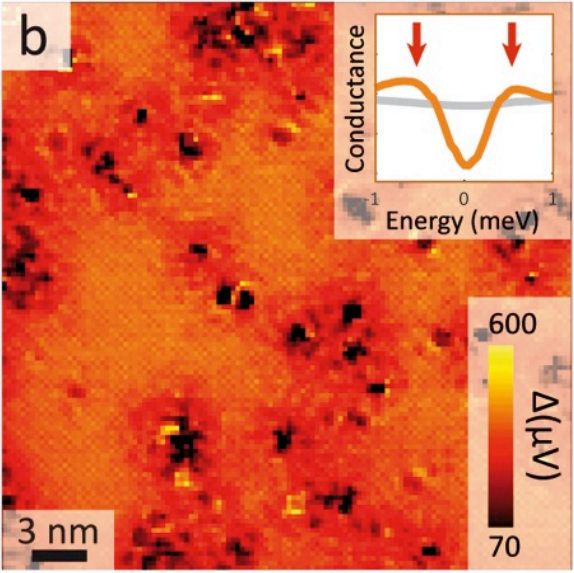
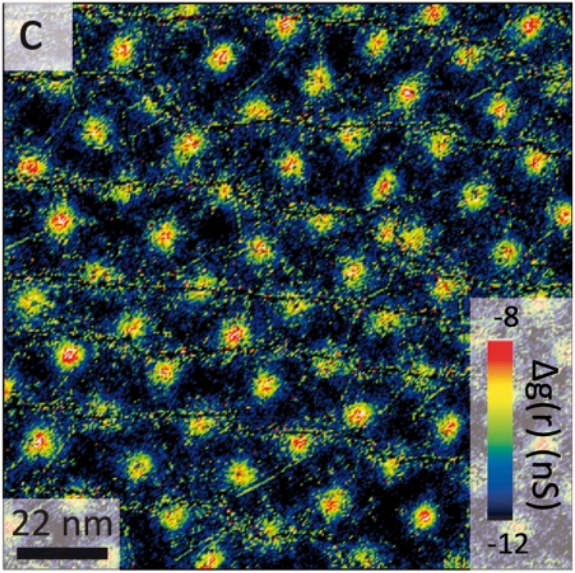

**Fig 3**

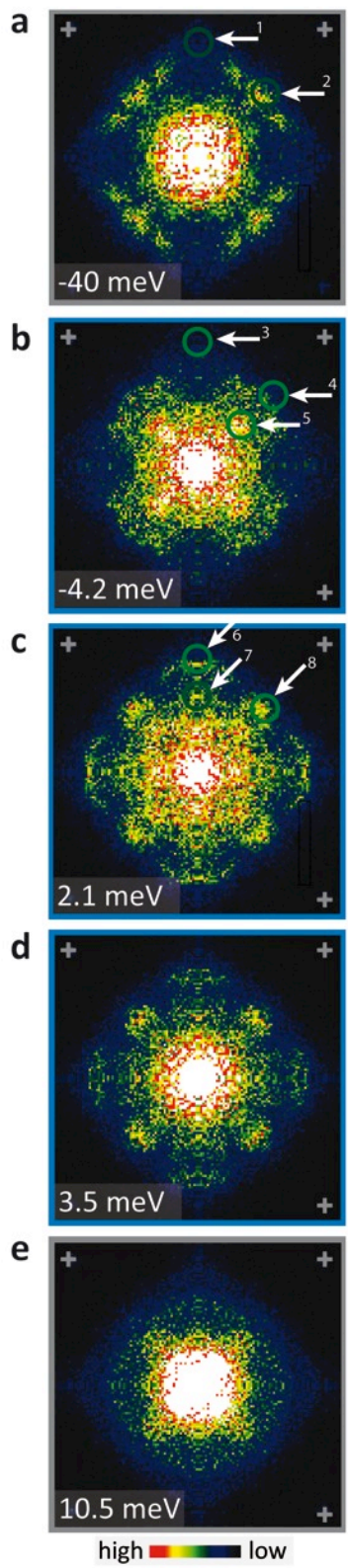
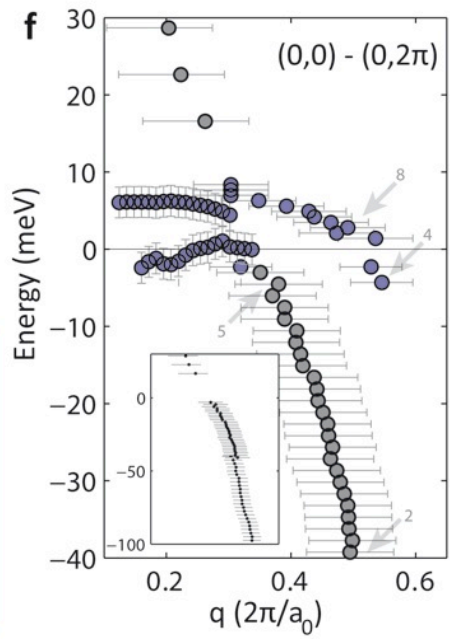
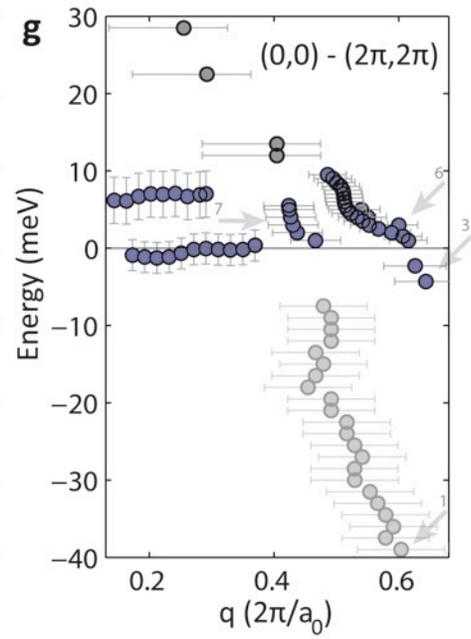
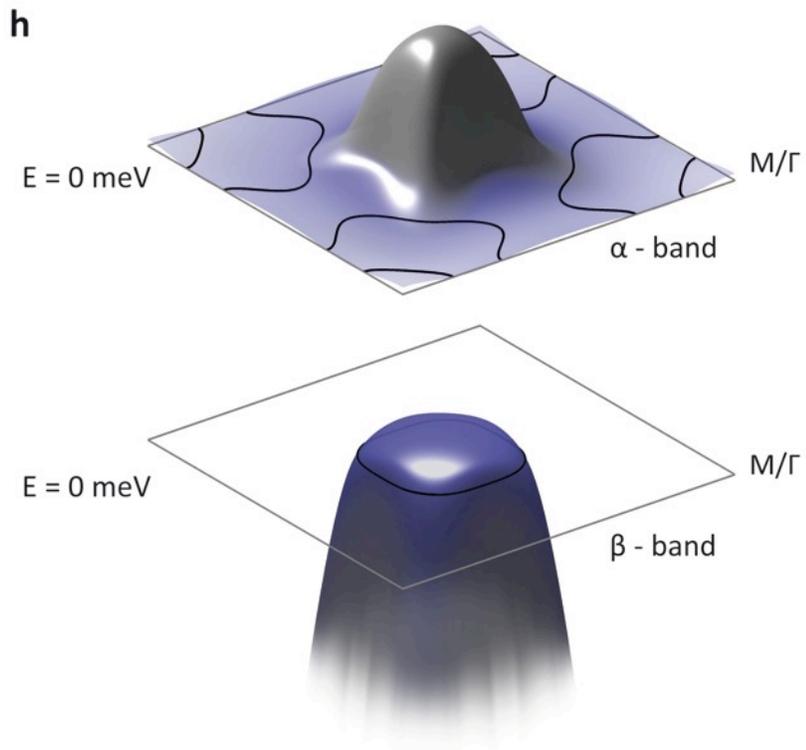

**Fig 4**

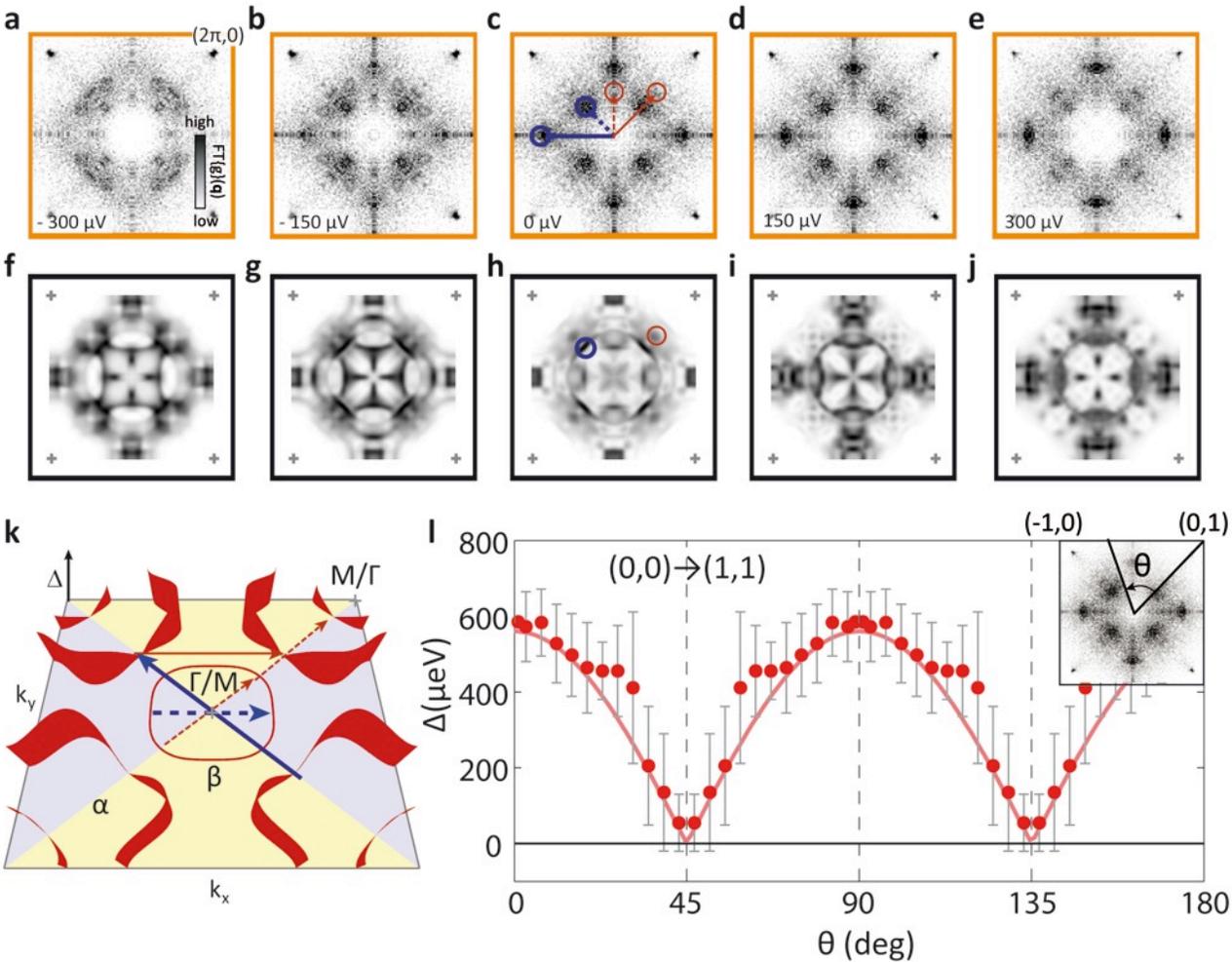

Supporting information for

# Imaging Cooper Pairing of Heavy Fermions in CeCoIn$_5$


M. P. Allan[§], F. Massee[§], D.K Morr, J. van Dyke, A.W. Rost, A. P. Mackenzie, C. Petrovic and J. C. Davis


## (I) Light band, heavy fermions, and Bogoliubov QPI

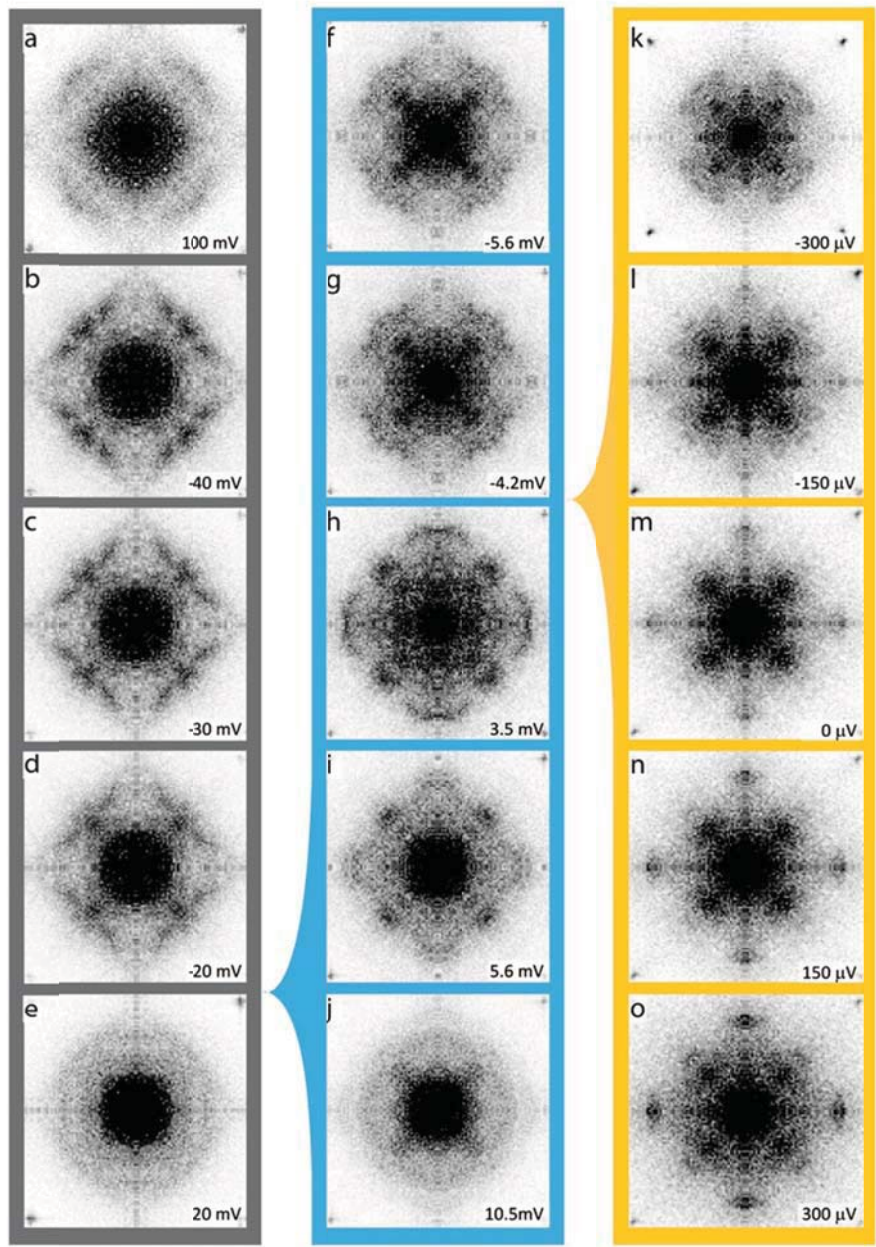

Fig. S1. Light (a-e), hybridized heavy-band (f-j) and superconducting Bogoliubov (k-o) quasiparticle scattering interference patterns, i.e. g($q$,E). The raw data were symmetrized to increase signal-to-noise.



To explore the complete physics of heavy fermion superconductivity in CeCoIn$_5$, we conducted spectroscopic imaging STM experiments on very different energy scales: High energy scales (± 100 meV) to measure the light bands, smaller scales (± 15meV) to reveal the hybridization of heavy bands, and very small energy scales (± 500 $\mu$eV) for the Bogoliubov quasiparticle interference (QPI) patterns specific to the superconductivity. Figure S1 shows some representative g($q$,E) layers, supplementary movies movieS1.avi and movieS2.avi show the evolution of the QPI patterns with energy (c.f. Section V)

## (II) Comparison between QPI data and theoretical simulations

In this section we describe how we generate and compare the theoretical simulations for the quasiparticle interference response in CeCoIn$_5$. In the Kondo screened state, hybridization between the light conduction band with dispersion $\varepsilon_\mathbf{k}$ and the heavy band with dispersion $\chi_\mathbf{k}$ leads to two new bands with energy dispersion

$$E_\mathbf{k}^{\alpha,\beta} = \frac{\varepsilon_\mathbf{k} + \chi_\mathbf{k}}{2} \pm \sqrt{\left(\frac{\varepsilon_\mathbf{k} + \chi_\mathbf{k}}{2}\right)^2 + V(\mathbf{k})^2} \tag{S1}$$

where $V(\mathbf{k}) = \bar{V}(\mathbf{k})r_0$ is the momentum dependent hybridization arising from coupling of the magnetic moment to neighboring conduction band sites. $\bar{V}(\mathbf{k})$ is the bare hybridization, and $r_0$ accounts for the renormalization of the bare hybridization due charge fluctuations. Within the slave-boson approach to the Anderson model, the effects of charge fluctuations are accounted for through the expectation value of the slave boson, $r_0$.

The energy dispersions presented in Eqs. (S1) are determined by requiring that the resulting simulated QPI spectrum, $\boldsymbol{g}_{th}(q,E)$, computed from Eqs.(8) and (9) in Ref. [1], reproduces the experimentally observed one. For the calculation of the heavy fermion QPI simulations shown in the manuscript we use U$_f$ /U$_c$=0.20 and U$_{fc}$ /U$_c$=0.265 . Here, $U_c$ and $U_f = U_f^0 r_0^2$ are the (renormalized) scattering potentials for intraband scattering in the light and heavy bands, respectively, while $U_{cf} = U_{fc} = U_{cf}^0 r_0$ is the renormalized scattering potential for interband scattering between the light and heavy bands. For the energy dispersions we use the form

$$\varepsilon_\mathbf{k} = -\mu_c - 2t_{c1}\left(\cos(k_x) + \cos(k_y)\right) - 4t_{c2}\cos(k_x)\cos(k_y) \\ - 2t_{f3}\left(\cos(2k_x) + \cos(2k_y)\right) \tag{S2a}$$

$$\chi_\mathbf{k} = -\mu_f - 2t_{f1}\left(\cos(k_x) + \cos(k_y)\right) - 4t_{f2}\cos(k_x)\cos(k_y) \\ -2t_{f3}\left(\cos(2k_x) + \cos(2k_y)\right) \\ - 4t_{f4}[\cos(2k_x)\cos(k_y) + \cos(k_x)\cos(2k_y)] \\ -4t_{f5}\cos(2k_x)\cos(2k_y) - 4t_{f6}\cos(3k_x)\cos(3k_y) \\ - 2t_{f7}\left(\cos(3k_x) + \cos(3k_y)\right) \tag{S2b}$$



To reproduce the experimental results (Figs S2, S3), we chose the following parameters $t_{c1} = -50.0$ meV, $t_{c2} = -13.4$ meV, $t_{c3} = 16.7$ meV, $\mu_c = 151.5$ meV, $t_{f1} = -1.0$ meV, $t_{f2} = -0.45$ meV, $t_{f3} = -0.7$ meV, $t_{f4} = 0$, $t_{f5} = 0.125$ meV, $t_{f6} = 0$, $t_{f7} = 0.15$ meV, and $\mu_f = -0.5$ meV. Moreover, to describe the experimental QPI dispersion, we find that it is necessary to introduce a momentum dependent hybridization of the form

$$V(\mathbf{k}) = V_0 + V_1 \left[\sin(k_x)\sin(k_y)\right]^2 \tag{S3}$$

where $V_0 = 3.0$ meV and $V_1 = 7.0$ meV. A comparison of the theoretically computed QPI spectrum along $q_y = 0$ and $q_y = q_x$ are shown in Fig. S2(a) and (b).

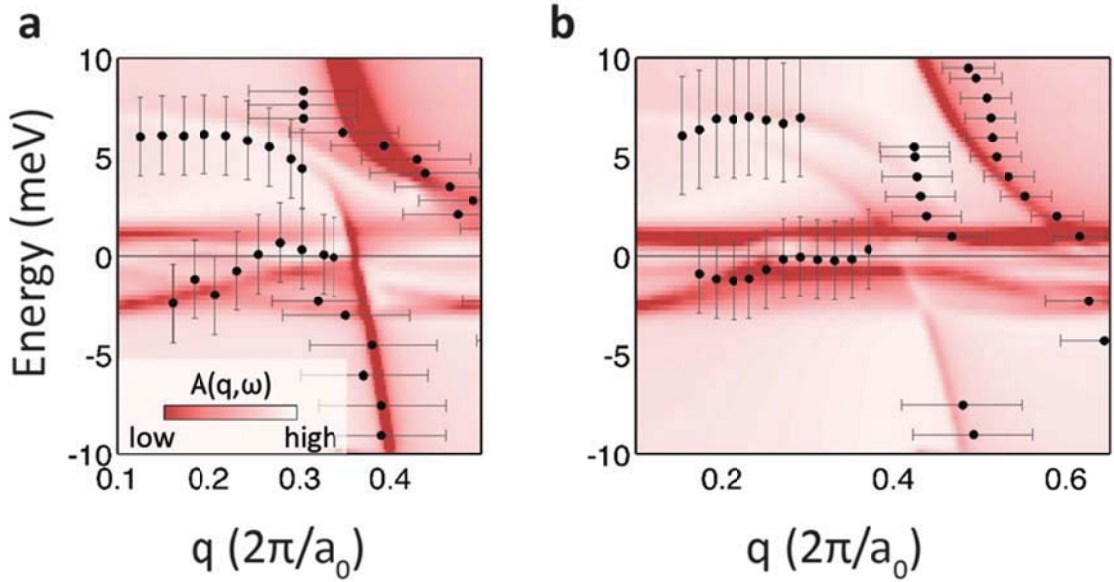

Fig. S2. Comparison between the light and heavy fermion band structure in the experimental data and that in the $g_{th}(|\mathbf{q}|,E)$ simulation for the (0,1) (a) and (1,1) (b) directions. The black dots mark the positions of maxima extracted from measured $g(\mathbf{q},E)$ layers and cuts as shown in Figure 3 in the main text; the false-color plot in the background shows the scattering intensities from the theoretical QPI calculations of $g_{th}(\mathbf{q},E)$.



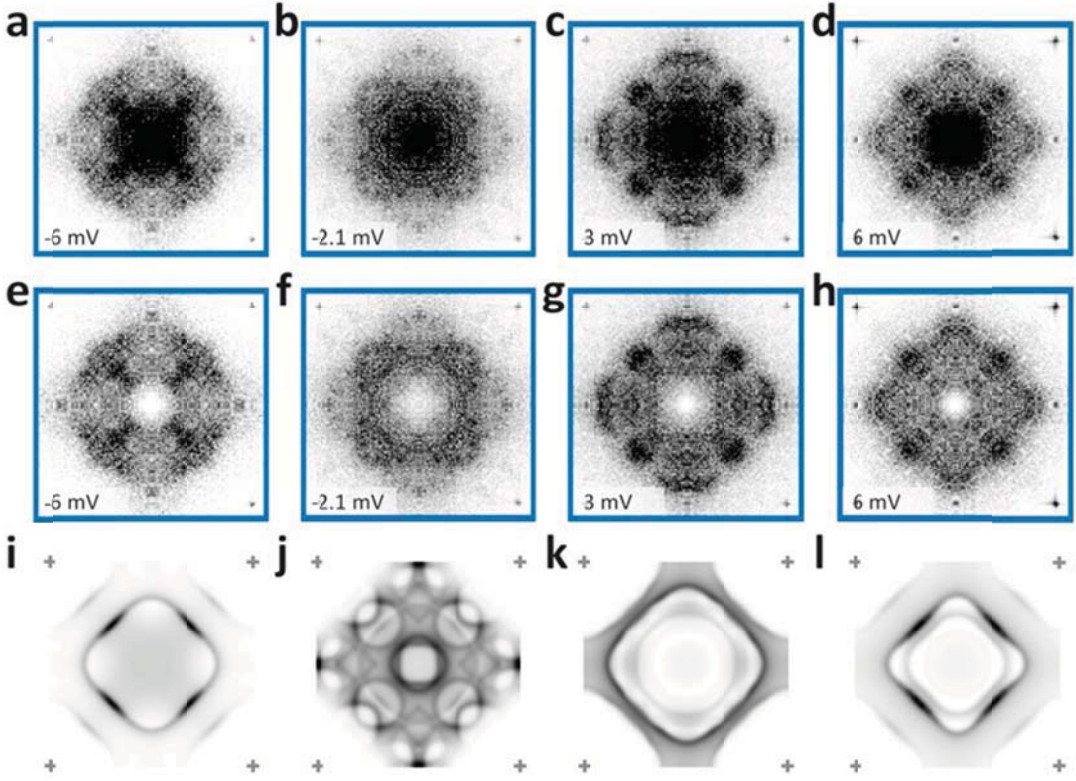

Fig. S3. Light and heavy fermion band structure comparison between experimental data (**a-h**) and simulation (**i-l**). The top row is the symmetrized data, while in the middle row the central core is suppressed using a Gaussian to allow for better comparison to the simulations (a-h treated as described in Section V).

In Fig. S3 we present a comparison of the layers of constant energy, $g_t(q, \omega = const)$ of the experimental QPI data, Fig.S3(a)-(h), along with the theoretical QPI simulations, Fig.S3(i)-(l). The comparisons shown in Figs S2 and S3 demonstrate the good agreement between the experimental and theoretical QPI spectra in the energy range outside the superconducting gap.

In presenting the theoretical results of Fig.S3(i) – (l), we accounted for two experimental effects, as demonstrated in Fig.S4. First, we applied a low-pass filter to the theoretical QPI spectra to simulate the finite q-resolution of the experimental data [see Fig.S4(b)]. We then apply the repeated zone scheme [Fig.S4(c)]. Finally, the experimental measurement of the density of states at 4 points within a unit cell implies that a structure factor

$$S(\mathbf{q}) = \frac{1}{2}\sqrt{\left[1 + \cos\left(\frac{q_x}{2}\right)\right]\left[1 + \cos\left(\frac{q_y}{2}\right)\right]} \ . \tag{S4}$$

should be applied to the theoretical data, leading to a suppression of the high-q intensity [Fig.S4(d)].



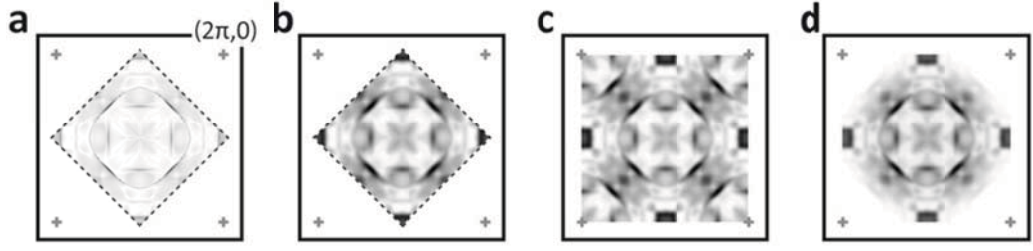

Fig. S4. Comparison between g(**q**,E) simulations and data. First we apply a low-pass filter to the original simulation shown in (a) to simulate the finite q-resolution of the experimental data, and obtain (b). We then apply the repeated zone scheme (c) to fill the full experimentally accessed reciprocal space. To simulate the rapid attenuation of larger q-vectors that is generally observed in QPI data, and that could stem from a structure factor $S(\mathbf{q})$ (see text) we suppress the high-q intensity in the simulations, resulting in (d).

Next, we explore the energy range where superconductivity occurs. In the superconducting state, the strong Fermi surface mismatch between the α– and β-bands implies that only the predominant pairing terms are of the form $\langle \alpha^+_{k,\uparrow}, \alpha^+_{-k,\downarrow} \rangle$ and $\langle \beta^+_{k,\uparrow}, \beta^+_{-k,\downarrow} \rangle$. As a result, the energy dispersion of the α– and β-bands are modified to yield

$$\Omega^{\alpha,\beta}_{\mathbf{k}} = \sqrt{\left(E^{\alpha,\beta}_{\mathbf{k}}\right)^2 + \left(\Delta^{\alpha,\beta}_{\mathbf{k}}\right)^2}. \tag{S5}$$

In Fig. S5, we present a comparison of experimental Bogoliubov-QPI (BQPI) g(**q**,E) for three different energies inside the superconducting gap (Fig. S5a-c) along with theoretical BQPI spectra for the same energies and two different symmetries of the superconducting order parameter. The effects of finite **q**-resolution and of the structure function, were again taken into account ( Fig.S4 ). The theoretical BQPI simulations in the Born approximation for potential scattering are given by

$$\boldsymbol{g}_{th}(q,\omega) = 2\frac{\pi e}{\hbar} N_t \sum_{i,j=1,2} \left[\hat{t}\widehat{N}(q,\omega)\hat{t}\right]_{ij} \tag{S6a}$$

$$\widehat{N}(q,\omega) = -\frac{1}{\pi}\text{Im}\left[\int \frac{d^2 k}{(2\pi)^2} \hat{G}(k,\omega)\widehat{U}\hat{G}(k+q,\omega)\right] \tag{S6b}$$

with

$$\hat{t} = \begin{pmatrix} -t_c & 0 & 0 & 0 \\ 0 & -t_f & 0 & 0 \\ 0 & 0 & t_c & 0 \\ 0 & 0 & 0 & t_f \end{pmatrix} \tag{S7}$$

and



$$\hat{U} = \begin{pmatrix} U_c & U_{cf} & 0 & 0 \\ U_{fc} & U_{ff} & 0 & 0 \\ 0 & 0 & -U_c & -U_{cf} \\ 0 & 0 & -U_{fc} & -U_{ff} \end{pmatrix}. \tag{S8}$$

Here, $N_t$ is the density of states in the STM tip, $t_c, t_f$ are the amplitudes for electron tunneling from the STM tip into the light and heavy bands, respectively, and, $\hat{G}(\mathbf{k},\omega)$ is the Fourier transform of the Green's function matrix (in Nambu notation)

$$\hat{G}(\mathbf{k},\tau) = -\langle T_\tau \Psi_k(\tau) \Psi_k^\dagger(0) \rangle \tag{S9}$$

where we defined the spinor

$$\Psi_k^\dagger = \left( c_{k,\uparrow}^\dagger, f_{k,\uparrow}^\dagger, c_{-k,\downarrow}, f_{-k,\downarrow} \right) \tag{S10}$$

and $c_{k,\sigma}^\dagger, f_{k,\sigma}^\dagger$ creates an electron in the light conduction and heavy moment bands, respectively.

We then impose a superconducting gap with $d_{x^2-y^2}$ symmetry for both the α– and β-bands

$$\Delta_k^{\alpha,\beta} = \frac{\Delta_0^{\alpha,\beta}}{2} (\cos k_x - \cos k_y) \tag{S11}$$

where $\Delta_0^\alpha$= 1.0 meV, $\Delta_0^\beta$= –0.20 meV. Note that for this value of $\Delta_0^\beta$, the maximum superconducting gap on the Fermi surface of the β-band is approximately 50 μeV, and thus smaller than the experimental energy resolution. The expected QPI for this case are shown in Fig.S5d-f, while those obtained with a gap of $d_{xy}$ symmetry

$$\Delta_k^{\alpha,\beta} = \Delta_0^{\alpha,\beta} \sin k_x \sin k_y \tag{S12}$$

where $\Delta_0^\alpha$= 0.75 meV, $\Delta_0^\beta$= -0.1meV are shown in Fig.S5g-i. For the case of the $d_{xy}$ symmetry gap, we have adjusted the values of $\Delta_0^{\alpha,\beta}$ to match the maximum superconducting gaps for the $d_{x^2-y^2}$ symmetry case. Finally, in Fig. S5j-l, we present the theoretical $g_{th}(\mathbf{q},E)$ for a $d_{x^2-y^2}$ gap [see Eq.(S11)], but with a maximum gap in the $\beta$-band which is of equal magnitude as that in the $\alpha$-band ($\Delta_0^\alpha$= 1.0 meV, $\Delta_0^\beta$= -2.6 meV).

A comparison of the energetically equivalent experimental and theoretical QPI g(**q**,E) images provide strong evidence for the existence of a superconducting gap with $d_{x^2-y^2}$ symmetry, with a significantly smaller maximum gap in the $\beta$-band, as described by the values of $\Delta_0^\alpha$ and $\Delta_0^\beta$ given below Eq.(S11).

Finally, we note that our model reflects a sign change between the superconducting gaps of the α- and β-bands. We have implemented such a sign change because the pairing mechanism may be electronic in nature and therefore likely repulsive. Any repulsive pairing interaction (allowing for interband Cooper pair scattering) likely favors a sign shift between the superconducting gaps in the α- and β-bands, as assumed in our model.

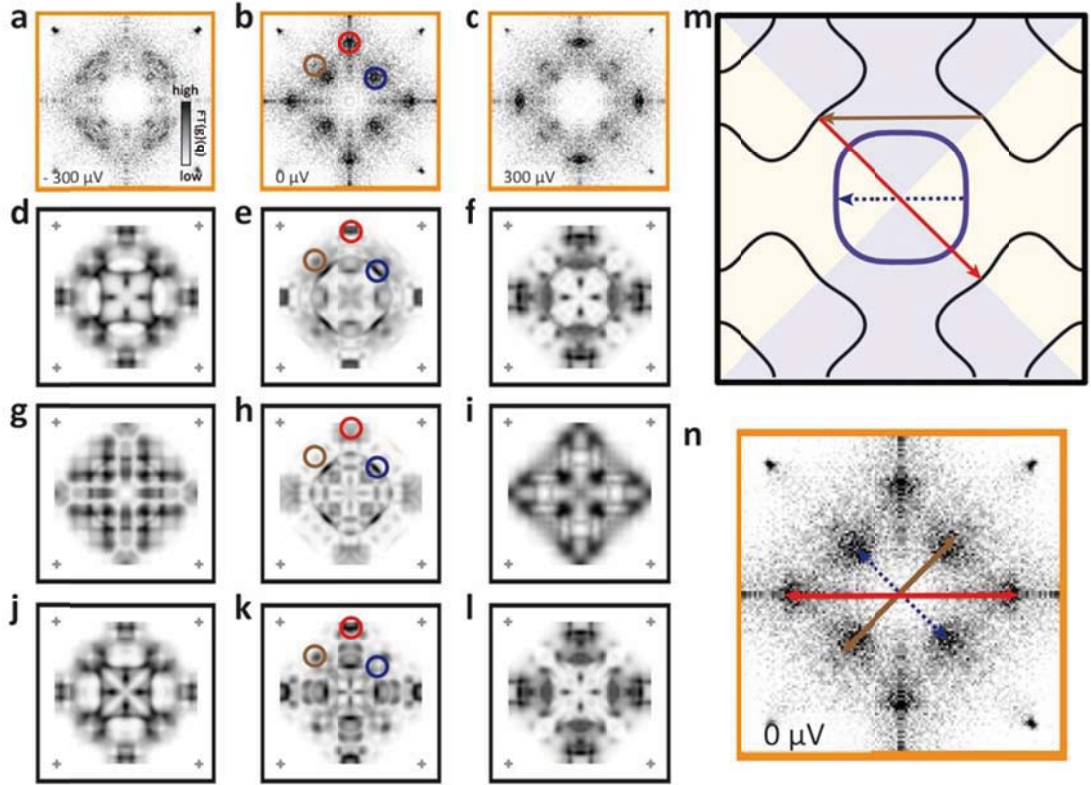

Fig. S5. Comparison between Bogoliubov g($q$,E) simulations and data for different superconducting gap symmetries. The experimental data (a-c) is compared to simulations using a $d_{x^2-y^2}$ gap symmetry (d-f), $d_{xy}$ gap symmetry (g-h) and a $d_{x^2-y^2}$ gap symmetry where the gaps on the α– and β-bands have equal magnitude (i-k). The red and brown circles highlight the intermodal scattering vectors, the blue circle highlights the strongest β-band scattering vector. Only the $d_{x^2-y^2}$ gap with small β-band gap reproduces all correctly. The important scattering vectors are also drawn schematically on the Fermi surface shown in (m), the corresponding features in the E=0mV experimental data are indicated in (n) (note that m and n are rotated by 45 degrees to one another).

### (III) Direct gap analysis

To obtain and estimate for the gap structure as a function of angle in momentum space, we use a procedure that is somewhat independent of the details of the Fermi surface: we integrate the total spectral weight g($q$,E) in each direction $|q|_\theta$ within the relevant $|q|$ range, i.e. $g(\theta, E) = \int_{q2}^{q1} g(|q|, E)$, to obtain the spectral weight as a function of angle and energy. Here we choose $q_1$ large enough to exclude small $|q|$ signal that stem from spatial heterogeneity and the small scattering vectors, and $q_2$ small enough to exclude the signal from the Bragg peak (Fig S6a). Figure S6 b,c show the results of such an analysis for g($q$,E) below and above $T_c$, respectively, Fig S6d shows the difference. Clearly, a gap in the spectral weight opens below $T_c$. This is emphasized in the particle hole symmetrized data, Fig. S6 e,f,g.

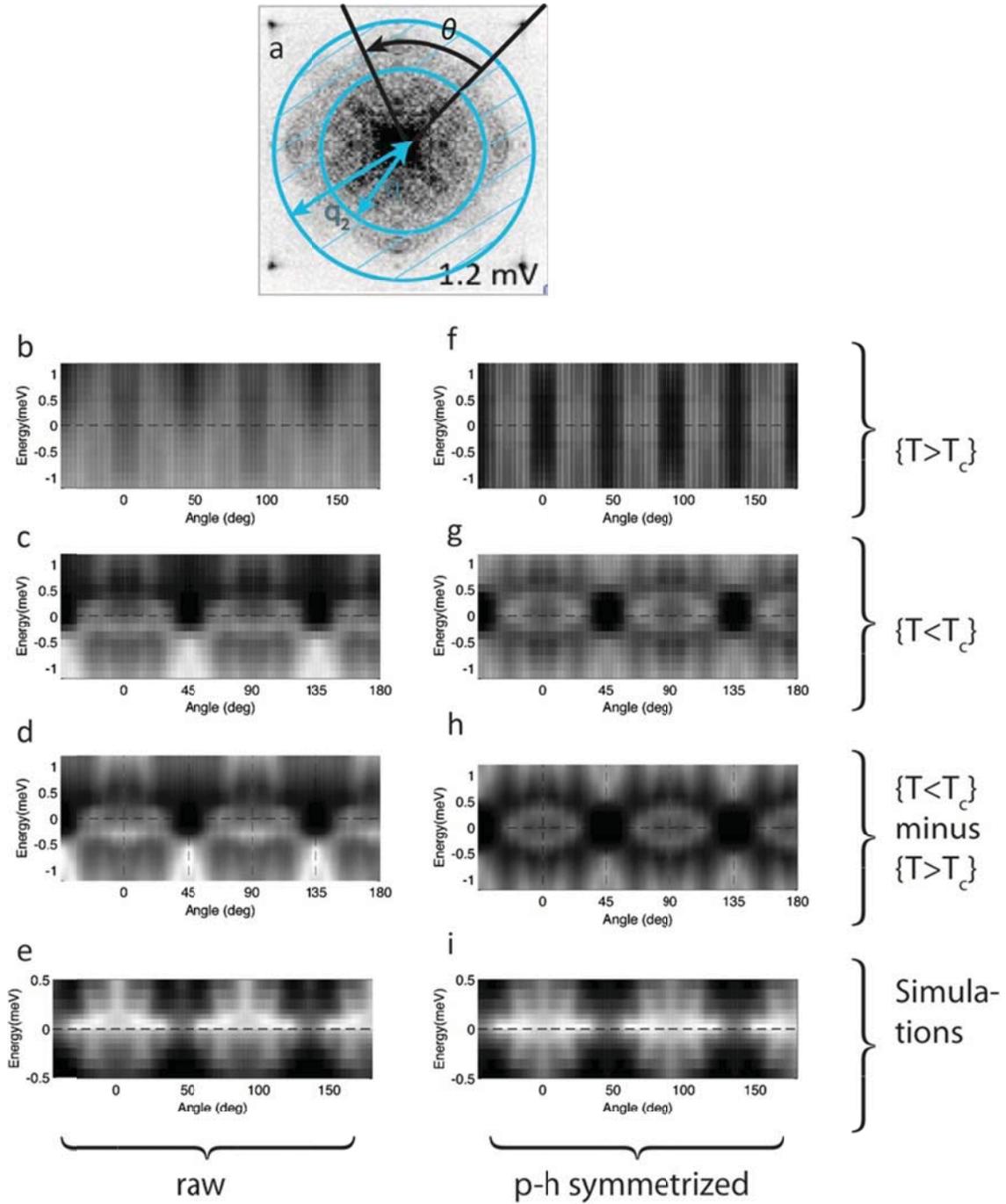

Figure S6 : Direct gap analysis. **a**, Layer of g(*q*,*E*=1.2meV) and the approximate range [$q_1$, $q_2$] over which we integrate. **b**, Result of the leading edge analysis for a dataset at *T*=2.5K>$T_c$. No gap in the spectral weight is visible. **c**, Same but for T=250mK<$T_c$, with a gap visible. **d**, Subtraction of data form above and below $T_c$, showing the influence of superconductivity on the spectral weight. **f-h**, Same as b-d but with particle hole symmetry enforced. **e,i**, Results achieved using same analysis scheme on the simulations described in Section I instead of data. All intensities are on a log scale.

To obtain the superconducting gap as a function of angle in reciprocal space, Δ(θ), as displayed in Figure 4 in the main text, we fit the image in Fig S6g at each angle θ with a gaussian plus a linear background, and extract the peak intensity positions. We note here that this procedure is not precisely valid for any complex bandstructure, but gives correct location of the nodes and the qualitative gap function quite independently of the exact Fermi surface. Importantly, the same procedure applied to the simulated QPI (Section II) yields very similar results (Fig. S6 d,h) providing strong support for its validity.

## (IV) Extracting the band structure from the data

The data points in Fig. 3f,g are extracted using two different methods. The light features that can accurately be tracked from the g(q,E) images are extracted directly from the g(q,E) layers – either by fitting line-cuts with a polynomial background plus a Gaussian or by following the points of largest intensity by eye (giving identical results). The very flat scattering features are extracted by fitting a polynomial background plus a Gaussian to the cuts along the two principal directions of the g(q,E) data-blocks. Fig. S7 shows such cuts where the intense flat QPI is clearly visible. The error bars in Fig. 3f,g of the main text indicate the width of the Gaussian or the estimated error in locating the maxima of intensity.

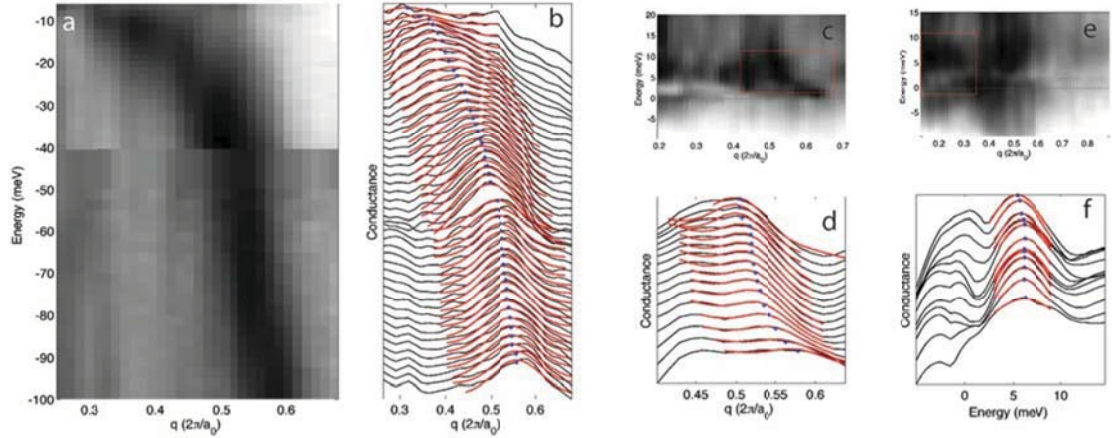

Figure S7 : Examples of the fitting procedure to extract the data-points in Fig 4.a,b form g(|q|,E) cuts. **a**, Cut of the g(|q|,E) intensity along the (0,0) – (0,1) direction. The light band is clearly visible with a hole like dispersion. **b**, fits with a constant background plus a Gaussian to the data to extract the positions of the maxima. **c-f**, Similar procedures for the heavy bands. The boxes in **c**, **e** indicate the range of **d,f**.

## (V) SI-STM procedure and setup effect

A well-known challenge of spectroscopic imaging STM is the so called setup effect, a consequence of the interplay between tip height and integrated electronic structure. All measurements reported here are performed in standard constant current mode, where the height of the tip is adjusted such that the current is identical for each location. In systems with weak dispersion such as CeCoIn$_5$ (conductance bands), the choice of current – and corresponding setup voltage – will strongly influence the observed QPI scattering pattern, see Fig. S8a-c. Details about the set up effect can be found e.g. in J. Chen's "Scanning tunneling microscopy" (Oxford Press, 2006) [2].

Different schemes exist to counter the setup effect in tunneling spectroscopy. One of them, named after Feenstra[3], is to normalize the differential conductance by the current divided by the voltage, (d$I$/d$V$)/($I$/$V$) and thereby cancelling the effect of tunneling matrix elements.

This technique is ideally suited for CeCoIn$_5$. We thus always measure simultaneously conductance $g(\mathbf{r},E)$ and current $I(\mathbf{r},E)$ maps, and apply the Feenstra technique to each spectrum $g(\mathbf{r}=const,E)$ individually, before the Fourier transform. Figure S8 shows how, upon doing so, the setup effect in the QPI data of CeCoIn$_5$ is nearly completely cancelled.

The g($q$,$E$) images shown throughout this work – except for the Bogoliubov g($q$,$E$) where the Feenstra scheme becomes more noise sensitive and the setup effect is limited – are therefore normalized using the Feenstra technique. The g($q$,$E$) are furthermore symmetrized along the crystallographic directions to suppress noise. (As Fig S9 shows, all reported key features are also clearly visible in the non-symmetrized g($q$,$E$)).

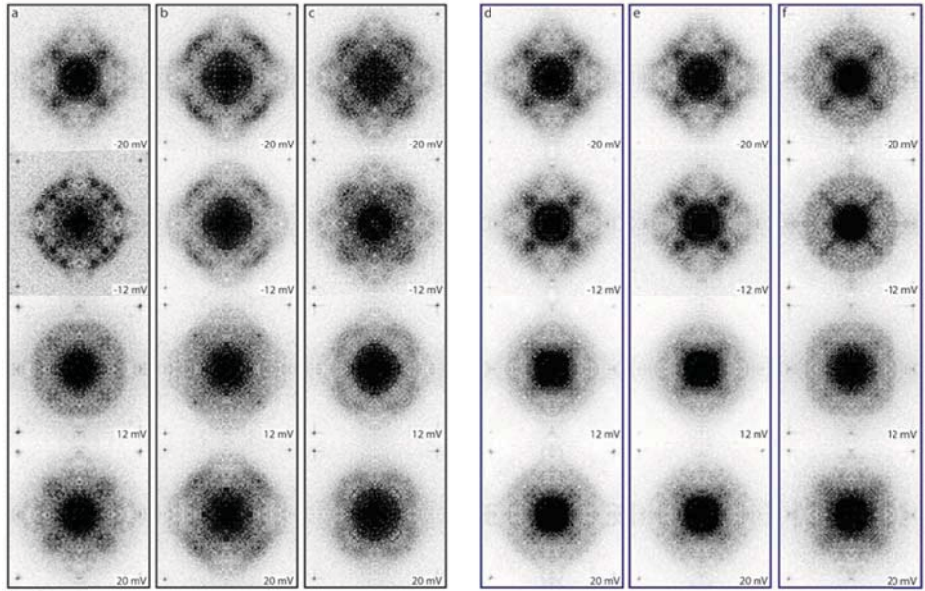

Figure S8: Setup effect and Feenstra normalization. **a-c**, Fourfold symmetrized Fourier transforms of g($r$,$E$) maps taken with different setup conditions at 0.25K (**a**) and 1.2K (**b, c**). **a**, Vs = -30mV, I = 800pA, **b**, Vs=-100mV, I=500pA, **c**, Vs=100mV, I=500pA. (**d-f**), Fourfold symmetrized Fourier transforms of g($r$,$E$)/(I($r$,$E$)/V where **d-f** corresponds to **a-c** respectively.

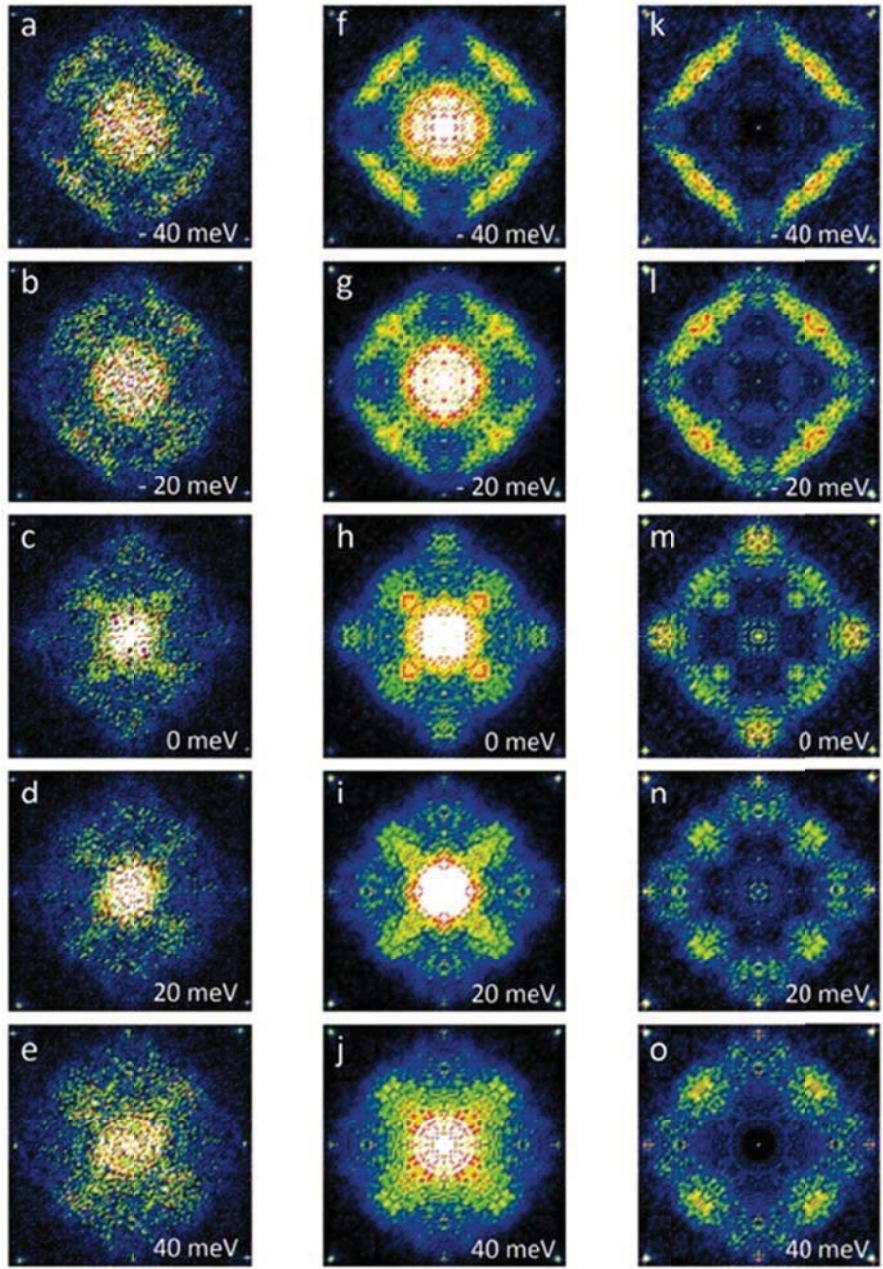

Figure S9: **a-e**, Raw QPI data, i.e. FFT($g(r,E)$/$I(r,E)$). **f-j**, Symmetrized QPI data as shown in the main text. **k-o**, Symmetrized QPI data with the core subtracted as e.g. in Ref [4].

---

[1] T. Yuan, J. Figgins, and D.K. Morr, Phys. Rev. B **86**, 035129 (2012).
[2] J. Chen, "Scanning tunneling microscopy" (Oxford Press, 2006)
[3] R. M. Feenstra *et al*. Surface Science **181**, 295-306 (1987)
[4] M.P. Allan et al. *Science* **336**, 563 (2012)